\documentclass[journal]{IEEEtran}
\IEEEoverridecommandlockouts
\usepackage{amsmath,amssymb}
\usepackage{subfigure}
\usepackage[mathscr]{eucal}
\usepackage{cite,verbatim}
\usepackage{epsfig}

\newcommand{\RR}{{\mathcal R}}
\newcommand{\Rdpc}{\RR_{DPC}}
\newcommand{\calX}{{\mathcal X}}
\newcommand{\calY}{{\mathcal Y}}
\newcommand{\CC}{{\mathcal C}}
\newcommand{\Mset}{{\mathcal M}}
\newcommand{\Cif}{\CC_{IFC}}
\newcommand{\Gif}{\CC_{GIFC}}
\DeclareMathOperator{\Exp}{{\mathbb E}} 
\newcommand{\abs}[1]{\lvert#1\rvert}
\newcommand{\var}[1]{Var(#1)}

\newtheorem{thm}{Theorem}[section]

\newtheorem{lem}[thm]{Lemma}
\newtheorem{cor}[thm]{Corollary}

\newtheorem{defn}{Definition}[section]

\begin{document}

\title{On the Capacity of Interference Channels with Degraded Message sets}
\author{Wei Wu, Sriram Vishwanath and Ari Arapostathis
\thanks{The authors are with Wireless Networking and Communications Group, 
Department of Electrical and Computer Engineering, The University of
Texas at Austin, Austin, TX 78712, USA (e-mail: \{wwu,sriram,ari\}@ece.utexas.edu). 
}
}
\maketitle

\begin{abstract}
This paper is motivated by a sensor network on a correlated field where nearby sensors share information, and can thus assist rather than interfere with one another. 
A special class of two-user Gaussian interference channels (IFCs) is considered where one of the two transmitters knows both the messages to be conveyed to the two receivers (called the IFC with degraded message sets).
Both achievability and converse arguments are provided for this scenario for a class of discrete memoryless channels with weak interference. For the case of the Gaussian weak interference channel with degraded message sets, optimality of Gaussian inputs is also shown, resulting in the capacity region of this channel.
\end{abstract}


\section{Introduction}

An interference channel (IFC), characterized by the channel $p(y_1,y_2|x_1,x_2)$, is one of the basic building blocks of many networks and is thus considered a fundamental problem in multi-user information theory. The capacity region of this channel remains an open problem, with some special cases such as the strong-interference case being solved. One of the fundamental difficulties faced while attacking the IFC capacity problem is that, unlike the broadcast channel, no transmit-side cooperation is possible.

Cooperation among transmitters can improve the rates achievable by employing joint encoding, i.e. \cite{Jindal:ISIT04_NodeCoop} \cite{Madsen:ISIT03_Coop}, \cite{Ng:ITW04_TXCoop}. 
As an example of this cooperation, Maric, Yates and Kramer consider the capacity of the strong interference channel with common information \cite{Maric:Allerton05IFC} and unidirectional cooperation \cite{Maric:ITA06}.

\begin{figure}[hbtp]
\begin{center}
	\subfigure[Data collection through IFC]{\label{fig:exampl_a} \includegraphics[scale=0.6]{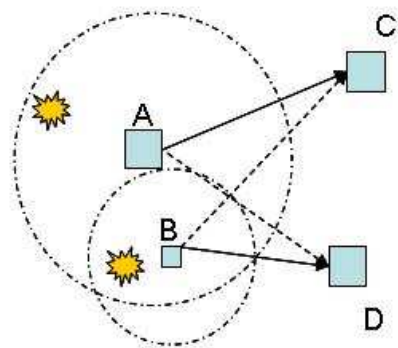}} \\
	\subfigure[Cascaded with degraded broadcast channel]{\label{fig:exampl_b} \includegraphics[scale=0.6]{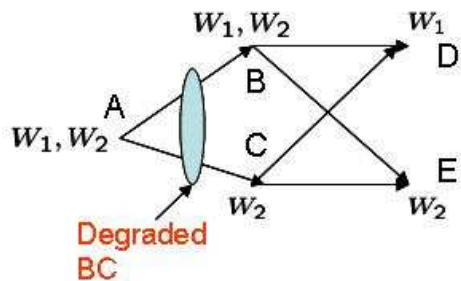}}
\caption{Two applications of interference channel with degraded message sets}
\end{center}
\label{fig:exampl}
\end{figure}

In this work, we consider a two-user IFC where we allow limited cooperation between the transmitters by means of permitting one of the transmitters to possess the message of the other, or in other words, the message sets are \emph{degraded}. Such a cooperative IFC is of interest on its own merit.
One example of this is shown in Figure~\ref{fig:exampl_a}, in which Sensor A has better sensing capability thus can detect both events while Sensor B can only detect one of them and the data need to be sent to different destinations. 
In this example, Sensor A can potentially improve Sensor B's transmission with the additional transmitter side information by cooperation. 
Another example is shown in Figure~\ref{fig:exampl_b}: $W_1$, $W_2$ is transmitted from $A$ to $D, E$ respectively;
if the intermediate nodes $B$, $C$ employ a decode-forward coding scheme with an inbuilt degraded broadcast channel, then $B$, $C$, $D$, $E$ form an IFC with degraded message sets, where $B$ can cooperate with $C$ in transmitting its message.
  
In this paper, we find an achievable region and outer bound on the capacity region of the discrete memoryless IFCs with degraded message sets. Specifically, for a class of  weak interference IFCs, this achievable region meets the outer bound giving us the capacity region for this class of channels. In the specific case of the Gaussian input distribution, we find Gaussian inputs to be optimal, resulting in the  region being characterized in closed form.

The rest of the paper is organized as follows. 
In Section~\ref{sec:prelim}, the basic definitions are presented and three types of weak interference are introduced. 
The main results are presented in Section~\ref{sec:main}, including the capacity region of a class of discrete memoryless weak interference channels (and Gaussian weak interference channels) with degraded message sets with numerical results of the Gaussian case shown in Section~\ref{sec:sim}. Detailed proofs for Section~\ref{sec:main} are given in Section~\ref{sec:proof}. 
Finally we conclude the paper with Section~\ref{sec:conclude}.    

\section{Notations and Preliminaries} \label{sec:prelim}
\subsection{Channel model and definitions}
We adopt the following notational conventions. Random variables (RVs) will be noted by capital letters, while their realizations will be denoted by the respective lower case letters. 
$X_m^n$ denotes the random vector $(X_m, \ldots, X_n)$, and $X^n$ denotes the random vector $(X_1, \ldots, X_n)$. 

\begin{figure}[hbtp]
\centering
\includegraphics[width=3.5in]{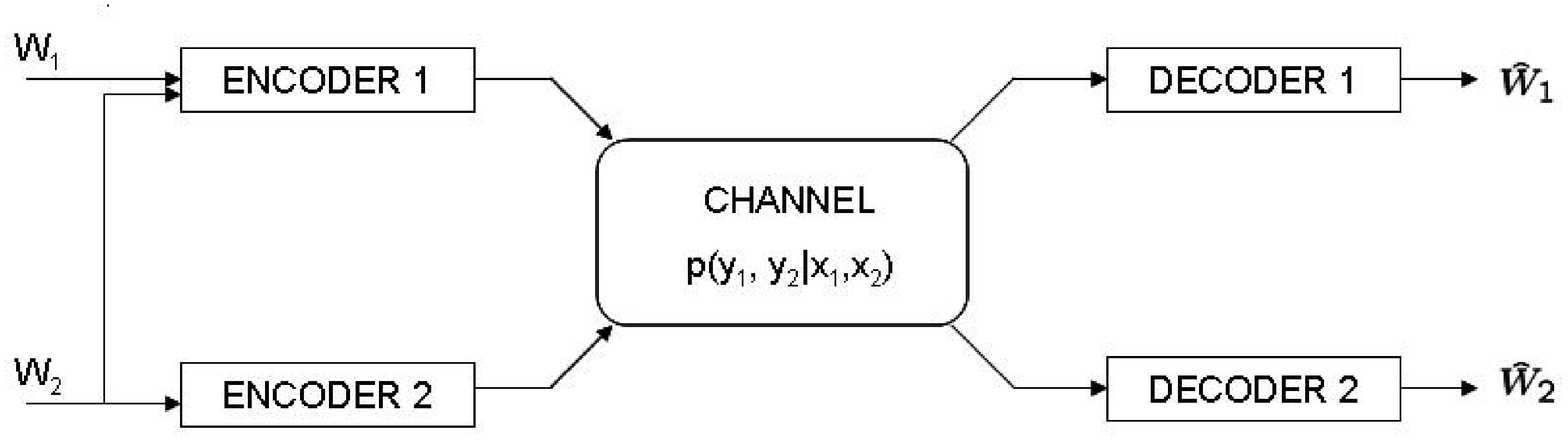}
\caption{The model of interference channel with degraded message sets}
\label{fig:IFC_DMS}
\end{figure}

A two-user \emph{interference channel} (IFC) $(\calX_1, \calX_2, \calY_1, \calY_2, P(y_1,y_2| x_1, x_2))$ is a channel with two input alphabets $\calX_1$, $\calX_2$, output alphabets $\calY_1$, $\calY_2$ and transition probability $P(y_1,y_2|x_1,x_2)$. 
It is assumed the channel is memoryless, namely
\begin{equation*}
P(y_1^n,y_2^n|x_1^n,x_2^n)=\prod_{i=1}^{n} P(y_{1,i},y_{2,i}|x_{1,i}, x_{2,i})\,.
\end{equation*}

Transmitter $t$ sends a message with $M_t$ bits, $W_t$, to receiver $t$ in $n$ channel uses at rate $R_t = M_t/n$ bits per use. 
A $(R_1,R_2,n, P_{e,1}, P_{e,2})$ code is defined as 
any code achieving the rate pair $(R_1, R_2)$ with block size $n$ and 
decoding error probability $P_{e,t}^{(n)}$, $t=1,2$. 
The \emph{capacity region} $\Cif$ is the closure of the set of rate pairs $(R_1, R_2)$, for which the receivers can decode their messages with error probability $P_{e,t}^{(n)} \to 0$ for $t=1,2$ as the block size $n \to +\infty$.

In the classic IFC described above, each transmitter has its own message set $\Mset_t = \{W_t\}$, where $W_t \in \{1, 2, \ldots, 2^{N R_t}\}$ denotes the private message to the receiver $t$.
In the IFC with common information, which is proposed recently by Maric, Yates and Kramer in \cite{Maric:Allerton05IFC}, each transmitter has not only its own private message $W_t$, but also the common message $W_0$ shared by all the transmitters. Thus the message set for transmitter $t$ is $\Mset=\{W_0, W_t\}$. 

In this paper, we consider the interference channel \emph{with degraded message sets} (IFC-DMS). For a two-user IFC with the degraded message sets, the message set of one transmitter is a strict subset of the other.
For example, Figure~\ref{fig:IFC_DMS} corresponds to the message sets,
\begin{equation}\label{eq:msgset_1}
\{W_2\}=\Mset_2 \subset \Mset_1=\{W_1, W_2\} \,,
\end{equation}
for which the capacity region is denoted as $\Cif^{T_1}$ regarding to that transmitter 1 knows both messages.
On the other hand, $\Cif^{T_2}$ denotes the capacity region of IFC with transmitter 2 knowing both messages, namely
\begin{equation}\label{eq:msgset_2}
\{W_1\}=\Mset_1 \subset \Mset_2=\{W_1, W_2\} \,.
\end{equation}

In recent work (e.g. \cite{Maric:ITA06}), the IFC-DMS has also been referred to as \emph{interference channel with unidirectional cooperation} - one transmitter knows the other's message and thus can enhance the achievable rate region.
The definition of degradedness of message sets can be further generalized to $K$-user IFC: the message sets are degraded if there exists a permutation $\{\sigma_k, k=1,\ldots,K\}$ of $\{1,2, \ldots,K\}$, such that 
\begin{equation*}
\Mset_{\sigma_1} \subset \ldots \subset \Mset_{\sigma_K}\,.
\end{equation*}   

In general, the capacity region of an IFC is an open problem and only known for certain classes of IFCs which include the so-called \emph{strong interference} channels, which satisfy
\begin{equation}\label{eq:strongIFC}
\begin{split}
I(X_1; Y_1|X_2) &\leq I(X_1; Y_2|X_2) \\
I(X_2; Y_2|X_1) &\leq I(X_2; Y_1|X_1) \,,
\end{split}
\end{equation}
for all product distributions on the inputs  $X_1$ and $X_2$. 
The capacity region in this case coinsides with the capacity region of compound IFC which is the union of two compound multiple access channels (MACs) determined by Ahlswede \cite{Ahlswede:strongIFC}.
Under a strong interference assumption that is (slightly) different from \eqref{eq:strongIFC}, given by
\begin{equation}\label{eq:strongIFC_common}
\begin{split}
I(X_1; Y_1|X_2, U) &\leq I(X_1; Y_2|X_2, U) \\
I(X_2; Y_2|X_1, U) &\leq I(X_2; Y_1|X_1, U) \,,
\end{split}
\end{equation}
for all joint distributions $P(u,x_1,x_2,y_1,y_2)$ that factor as $P(u)P(x_1|u)P(x_2|u)P(y_1,y_2|x_1,x_2)$,
Maric, Yates and Kramer find the capacity region of strong IFCs with common information \cite{Maric:Allerton05IFC} and with degraded message sets \cite{Maric:ITA06}. 

In this paper, we study three different notions of \emph{weak interference} and investigate the capacity region for this channel with degraded message sets. These three notions are extensions of the concepts of \emph{stochastically degraded}, \emph{less noisy} and \emph{more capable} from broadcast channel literature\cite{ElGamal:IT79BC}.
\begin{defn}\label{defn:typeA}
The interference channel is said to be \emph{type A weak interference channel} if 
\begin{equation} \label{eq:typeA}
I(X_1; Y_2|X_2) \leq I(X_1; Y_1|X_2) 
\end{equation}
or,
\begin{equation}
I(X_2; Y_1|X_1) \leq I(X_2; Y_2|X_1)\,.
\end{equation}
\end{defn}

\begin{defn}\label{defn:typeB}
The interference channel is said to be \emph{type B weak interference channel} if, for any $U$, $U \to (X_1, X_2) \to (Y_1, Y_2)$ forms a Markov chain, it satisfies 
\begin{equation} \label{eq:typeB}
I(U, X_2; Y_2) \leq I(U, X_2; Y_1) 
\end{equation}
or,
\begin{equation}
I(U, X_1; Y_1) \leq I(U, X_1; Y_2)\,.
\end{equation}
\end{defn}

\begin{defn}\label{defn:typeC}
The interference channel is said to be \emph{type C weak interference channel} if there exists a probability transition matrix $q_1(y_2|x_2,y_1)$ such that
\begin{equation} \label{eq:typeC}
p(y_2|x_1,x_2)=\sum_{y_1} p(y_1|x_1,x_2)q_1(y_2|x_2,y_1)\,,
\end{equation}
or if there exists a probability transition matrix $q_2(y_1|x_1, y_2)$ such that
\begin{equation}
p(y_1|x_1,x_2)=\sum_{y_1} p(y_2|x_1,x_2)q_2(y_1|x_1,y_2)\,.
\end{equation}
\end{defn}

Note that if the channel input $X_2$ of IFC is fixed, then the channel between $X_1$, $Y_1$, $Y_2$ forms a broadcast channel. 
For type C weak interference satisfying \eqref{eq:typeC}, this broadcast channel is \emph{stochastically degraded}.
This broadcast channel is \emph{less noisy} for type B weak interference satisfying \eqref{eq:typeB}, and \emph{more capable} for type A weak interference satisfying \eqref{eq:typeA} \cite{ElGamal:IT79BC}. 
Note that the notion of ``more capable'' is strictly weaker than that of ``less noisy'', which is strictly weaker than ``degraded''. 

In this paper, we will establish the capacity region of the interference channel with degraded message sets in Figure~\ref{fig:IFC_DMS} and type C weak interference.

\subsection{Gaussian interference channels}
One of our main interests in this paper is the Gaussian IFC, in which the alphabets of inputs and outputs are real numbers and the outputs are linear combinations of input signals and white Gaussian noise. 
The Gaussian IFC is defined as follows,
\begin{equation}
\begin{split}
Y_1 &=X_1+a X_2 + Z_1 \\
Y_2 &=b X_1 + X_2 + Z_2
\end{split}
\end{equation}
where $a$ and $b$ are real numbers and $Z_1$ and $Z_2$ are independent, zero-mean, unit-variance Gaussian random variables.
Furthermore, the transmitters are subject to average power constraints:
\begin{equation}
\lim_{N \to \infty} \frac{1}{N}\sum_{n=1}^{N} \Exp[X_{t n}^2] \leq P_t\,, \quad t=1,2\,.
\end{equation}

The capacity  region of the non-cooperative Gaussian IFC is currently characterized  for the cases when $a=b=0$ (trivial) or if $a^2 \geq 1$ and $b^2 \geq 1$, in which case the strong interference conditions in \eqref{eq:strongIFC} are satisfied. 
The capacity of IFC for strong interference is the set of $(R_1, R_2)$ satisfying \cite{Sato:IT81StrongIF,Han:IT81StrongIF}
\begin{align}
0 &\leq R_1 \leq \frac{1}{2}\log(1+P_1)\label{eq:IFC_R1}\\
0 &\leq R_2 \leq \frac{1}{2}\log(1+P_2)\label{eq:IFC_R2}\\
0 &\leq R_1+R_2 \leq \frac{1}{2}\log(P_1+ a^2 P_2+1)\\
0 &\leq R_1+R_2 \leq \frac{1}{2}\log(b^2 P_1+P_2+1)\,.
\end{align}  

In the regime when $0 \leq a^2 \leq 1$ or $0 \leq b^2 \leq 1$, the Gaussian IFCs belong to all three notions of weak interference.  Achievable rate regions \cite{Han:IT81StrongIF, Costa:IT85IFC, Sason:IT04IFC} and outer bounds \cite{Sato:IT77TwoUser, Carleial:IT83outbound, Costa:IT85IFC, Kramer:IT04outbound} are known for this scenario, but a characterization of the region is yet to be obtained.
A recent outer bound by Kramer in \cite{Kramer:IT04outbound} is given by $(R_1, R_2)$ satisfying \eqref{eq:IFC_R1}, \eqref{eq:IFC_R2}, and 
\begin{align}
R_1+R_2 & \leq \frac{1}{2}\log \Bigl[(P_1+a^2 P_2+1) \bigl(\frac{P_2+1}{\min(a^2,1) P_2+1} \bigr) \Bigr] \label{eq:sum_T2}\\
R_1+R_2 & \leq \frac{1}{2}\log \Bigl[(P_2+b^2 P_1+1) \bigl(\frac{P_1+1}{\min(b^2,1) P_1+1} \bigr) \Bigr] \label{eq:sum_T1}\,.
\end{align}

\begin{figure}[hbtp]
\begin{center}
	\subfigure[The Gaussian IFC with degraded message sets]{\label{fig:GIFC_DMS} \includegraphics[scale=0.6]{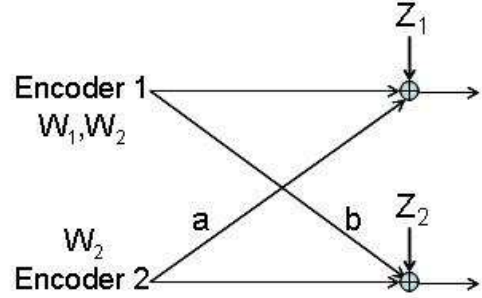}} \\
	\subfigure[The Gaussian MIMO broadcast channel]{\label{fig:GBC_bound} \includegraphics[scale=0.6]{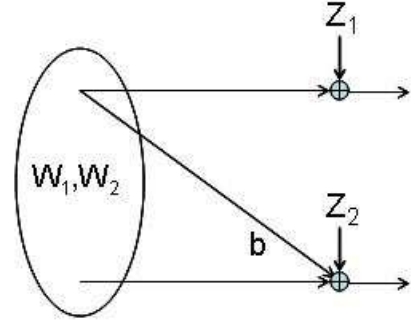}}
\caption{The Gaussian MIMO BC outer bound for IFC-DMS}
\end{center}
\label{fig:GIFC_BC}
\end{figure}

Let the capacity region of a Gaussian IFC with degraded message sets (i.e., Figure~\ref{fig:GIFC_DMS}) be denoted by $\Gif^{T_1}$. As a first step, we present an outer bound for $\Gif^{T_1}$.   

\subsection{Outer bound for Gaussian IFC-DMS}

First, we provide an ``intuitive framework'' for an outer bound on this channel. A formal outer bound follows this framework. 

A straightforward outer bound for Gaussian IFC with degraded message sets (see Figure~\ref{fig:GIFC_DMS}), $\Gif^{T_1}$, is the capacity region of the  Gaussian broadcast channel resulting from allowing full transmitter-side cooperation. 
One can make an even tighter outer bound by using the following arguments: 
\begin{enumerate}
\renewcommand{\theenumi}{\roman{enumi}}
\item Removing the interference link from encoder 2 to receiver 1 in Figure~\ref{fig:GIFC_DMS} does not enhance the overall capacity region because encoder 2 does not have any knowledge of message $W_1$ and  thus no cooperation can be induced to improve the transmission rate $R_1$; 
\item Now allowing  full cooperation between the two encoders provides us with an new broadcast channel shown in Figure~\ref{fig:GBC_bound}, which has two transmit antennas and one receive antenna at each receiver and an individual power constraint at each antenna.  
\end{enumerate}

Using the existing literature on the  capacity region of Gaussian multi-antenna (MIMO) BC channel \cite{Shamai:IT_BCCapacity}; dirty paper coding (DPC)\cite{Costa:IT84DPC} optimizes this outer bound \cite{Sriram:IT03GMBC} \cite{Tse:IT03GMBC}. If, in the dirty paper coding strategy, $W_1$ is encoded first and $W_2$ second, the rates achieved are given by:
\begin{multline}\label{eq:DPC_R12}
\Rdpc^{12}=\Bigl\{
(R_1, R_2): \quad \text{for}\quad 0 \leq \alpha \leq 1\,, \\
R_1 \leq \frac{1}{2}\log\bigl(1+ \alpha P_1 \bigr)\,, \\
R_2 \leq \frac{1}{2}\log\bigl(\frac{1+P_1 b^2+2\abs{b}\sqrt{(1-\alpha)P_1P_2}+P_{2}}{1+ b^2 \alpha P_1}\bigr)
\Bigr\}\,.
\end{multline} 
The DPC achievable region for the encoding sequence $W_2$, $W_1$ is 
\begin{multline} \label{eq:DPC_R21}
\Rdpc^{21}=\Bigl\{
(R_1, R_2): \quad \text{for}\quad 0 \leq \alpha \leq 1\,, \\
R_1 \leq \frac{1}{2}\log\bigl(\frac{1+P_1}{1+(1-\alpha)P_1} \bigr)\,, \\
R_2 \leq \frac{1}{2}\log\bigl(1+P_1 b^2+2\abs{b}\sqrt{(1-\alpha)P_1P_2}+P_{2}\bigr)
\Bigr\}\,.
\end{multline}

It is not hard to see that the capacity region of Gaussian IFC with degraded message set in Figure~\ref{fig:GIFC_DMS} is contained in the union of these two regions 
\begin{equation*}
\Gif^{T_1} \subset \Rdpc^{12} \cup \Rdpc^{21}=\Rdpc^{21}\,.
\end{equation*}

 We show later in this paper that the capacity region $\Gif^{T_1}$ under weak interference is indeed equal to $\Rdpc^{12}$, and thus, even though the outer bound above is in general loose, it captures the intuition behind the optimal coding strategy for this channel. 

\section{Main results}\label{sec:main}
In this section, we will first give our inner and outer bounds for the discrete-memoryless weak interference channel with degraded message sets, and then specialize to Gaussian weak IFC-DMS and establish its capacity region. 

Define $\RR_{in}$ to be the set of all rate $(R_1, R_2)$ such that 
\begin{equation}
\begin{split}
R_1 &\leq I(V; Y_1)-I(V; U, X_2)\\
R_2 &\leq I(U, X_2; Y_2)\,,
\end{split}
\end{equation}
for the probability distribution $p(x_1, x_2, u, v, y_1, y_2)$ that factors as 
\begin{equation*}
p(u, x_2)p(v|u, x_2)p(x_2|v)p(y_1,y_2|x_1,x_2)\,.
\end{equation*} 

The following theorem gives the achievable region of IFC with transmitter 1 knowing both messages as in Figure~\ref{fig:IFC_DMS} using the Gel'fand-Pinsker coding scheme \cite{Gelfand:80noncausal}.
\begin{thm}\label{thm:achievable}
The capacity region of discrete memoryless interference channel with degraded message sets as \eqref{eq:msgset_1} satisfies
\begin{equation*}
\RR_{in} \subset \Cif^{T_1}\,.
\end{equation*}
\end{thm}

The proof is based on the Gel'fand-Pinsker coding scheme, in which $(U, X_2)$ is considered as the random parameters for the channel between $X_1$ and $Y_1$.
It is given in Section~\ref{sec:Proof_achieve}. 

The outer bound is stated next.
Define $\RR_{o}$ to be the set of all rate pairs $(R_1, R_2)$ such that
\begin{equation}\label{eq:bound_general}
\begin{split}
R_1 & \leq I(X_1; Y_1|X_2)\\
R_2 & \leq I(U, X_2; Y_2)\\ 
R_1+R_2 & \leq I(X_1; Y_1|U, X_2)+ I(U, X_2; Y_2)\,,
\end{split}
\end{equation}
for the probability distribution $p(x_1, x_2, u, y_1, y_2)$ that factors as 
\begin{equation*}
p(u, x_2)p(x_1|u)p(y_1,y_2|x_1,x_2)\,.
\end{equation*} 

\begin{thm}\label{thm:outer_general}
The capacity region of discrete memoryless interference channel with degraded message sets satisfies
\begin{equation*}
\Cif^{T_1} \subset \RR_{o}\,.
\end{equation*}
\end{thm}

The proof is provided in Section~\ref{sec:Proof_outer}.
Both Theorem~\ref{thm:achievable} and Theorem~\ref{thm:outer_general} hold for the general interference channels.
However, as we can see, there is a gap between the achievable region obtained in Theorem~\ref{thm:achievable} and the outer bound in Theorem~\ref{thm:outer_general}.

Next we put additional weak interference assumptions (type B and type C) and consider its capacity region.  
The key idea is to find a rate region, which is the achievable region under type B weak interference condition and the outer bound under type C weak interference condition, thus the capacity region of type C weak interference channels.

Define the rate region $\RR_{*}$ to be the set of all rate pairs $(R_1, R_2)$ such that
\begin{equation}\label{eq:R_star}
\begin{split}
R_1 & \leq I(X_1; Y_1|U, X_2) \\
R_2 & \leq I(U, X_2; Y_2)\,,  
\end{split}
\end{equation}
for the probability distribution $p(x_1, x_2, u, y_1, y_2)$ that factors as 
\begin{equation*}
p(u, x_2)p(x_1|u)p(y_1,y_2|x_1,x_2)\,.
\end{equation*} 

For type B weak interference channels, we have the following achievable region,
\begin{thm}\label{thm:innerB}
The capacity region of discrete memoryless type B weak interference channel with degraded message sets and that \eqref{eq:typeB} holds, satisfies
\begin{equation*}
\RR_{*} \subset \Cif^{T_1,B}  \,.
\end{equation*}
\end{thm}
The proof is based on the coding scheme for degraded broadcast channel and the sketch of the proof is provided in Section~\ref{sec:Proof_innerB} for completeness.
Note since type C weak interference channels satisfy \eqref{eq:typeB} also, thus 
\begin{equation}\label{eq:innerC}
\RR_{*} \subset \Cif^{T_1,C}
\end{equation}  
holds.

For type C weak interference channels, we can also show $\RR_{*}$ is also its outer bound. 

\begin{thm}\label{thm:capC}
The capacity region of discrete memoryless type C weak interference channel with degraded message sets and that \eqref{eq:typeC} holds, satisfies
\begin{equation*}
\Cif^{T_1,C} = \RR_{*}\,.
\end{equation*}
\end{thm}
The proof is given in Section~\ref{sec:Proof_capC}.

Next we will specialize the weak interference channel to the Gaussian case.  
Since Gaussian weak interference channels satisfy type C weak interference condition, the capacity region of Gaussian weak interference channel with degraded message sets can be established by Theorem~\ref{thm:capC} and proving optimality of Gaussian input.

\begin{thm}\label{thm:GIFC}
The capacity region of the Gaussian IFC with Gaussian inputs and transmitter $T_{1}$ knowing both messages, $\Gif^{T_1}$, when $\abs{b} \leq 1$, is the set of all rate pairs $(R_1, R_2)$ such that, for $0 \leq \alpha \leq 1$, 
\begin{align}
R_1 & \leq \frac{1}{2}\log\left(1+ \alpha P_1 \right)\label{eq:cap_R1}\\
R_2 & \leq \frac{1}{2} \log\left(1 + \frac{h \Sigma h^t}{1+ b^2 \alpha P_1}\right) \nonumber\\
 & = \frac{1}{2} \log\left(\frac{1+P_1 b^2+2\abs{b}\sqrt{(1-\alpha)P_1 P_2}+P_{2}}{1+ b^2 \alpha P_1} \right)  \label{eq:cap_R2}
\end{align}  
Here $h$ is the vector $[b ~1]$, and $\Sigma$ is a $2 \times 2$ covariance with diagonal elements equaling $(1-\alpha)P_1$ and $P_2$ respectively.
\end{thm}

As we have seen, the capacity region established in Theorem~\ref{thm:GIFC} is exactly equal to the dirty paper region in \eqref{eq:DPC_R12}, i.e.,
\begin{equation*}
\Gif^{T_1}=\Rdpc^{12}\,,
\end{equation*}
when the encoding sequence is $W_1$, $W_2$ for the Gaussian BC channel. 

By swapping the parameters of two transmitters, the capacity region $\Gif^{T_2}$ can be obtained as the following corollary. 
\begin{cor}\label{cor:GIFC}
The capacity region of the Gaussian IFC with Gaussian inputs and transmitter 2 knowing both messages, $\Gif^{T_2}$, when $\abs{a} \leq 1$, is the set of all rate pairs $(R_{1},R_{2})$ such that, for $0 \leq \beta \leq 1$,
\begin{equation}\label{eq:cap_T2}
\begin{split}
R_1 & \leq \frac{1}{2} \log\bigl(\frac{1+P_1 +2\abs{a}\sqrt{(1-\beta) P_1P_2}+a^{2}P_{2}}{1+ a^2 \beta P_2} \bigr)  \\
R_2 & \leq \frac{1}{2}\log\bigl(1+ \beta P_2 \bigr) \,.
\end{split}
\end{equation} 
\end{cor}

\section{numerical results}\label{sec:sim}
In this section, we compare the capacity region of Gaussian IFCs with degraded message sets with the outer bounds of the normal Gaussian IFCs and the rate regions of Gaussian BC channels by numerical results.

We consider the symmetric Gaussian interference channel with $P_1=P_2=6$ and $a^2=b^2=0.3$. 
The rate units are bits per channel use.
First we compare the capacity region of GIFC-DMS, $\Gif^{T_1}$ with the dirty paper coding regions $\Rdpc^{12}$ in \eqref{eq:DPC_R12}, $\Rdpc^{21}$ in \eqref{eq:DPC_R21} and the outer bound of GIFC in \eqref{eq:sum_T2} -- \eqref{eq:sum_T1}. 
As we can see, the capacity region of GIFC-DMS is strictly larger than the outer bound of GIFC, and the gap between these two shows the performance improvement by allowing encoders to cooperate partially.
The point when $\alpha=0$ corresponds to the full cooperation between two encoders to transmit message $W_2$.

\begin{figure}[hbtp]
\centering
\includegraphics[width=3in]{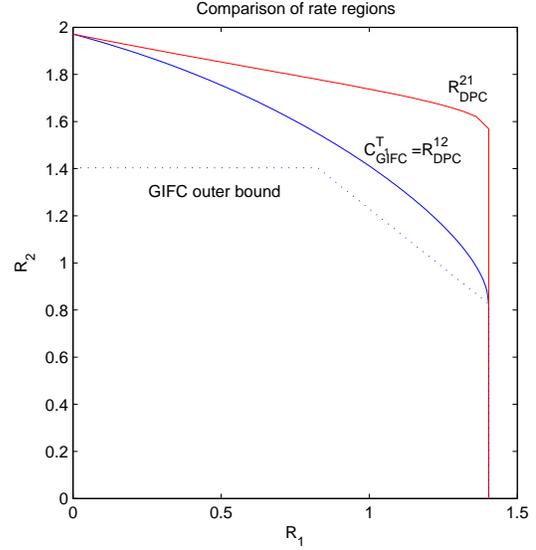}
\caption{The capacity region of Gaussian IFC-DMS with $P_1=P_2=6$, $a^2=b^2=0.3$ and achievable rate regions of Gaussian BC and IFC}
\label{fig:ifc_comp1}
\end{figure}

\begin{figure}[hbtp]
\centering
\includegraphics[width=3.5in]{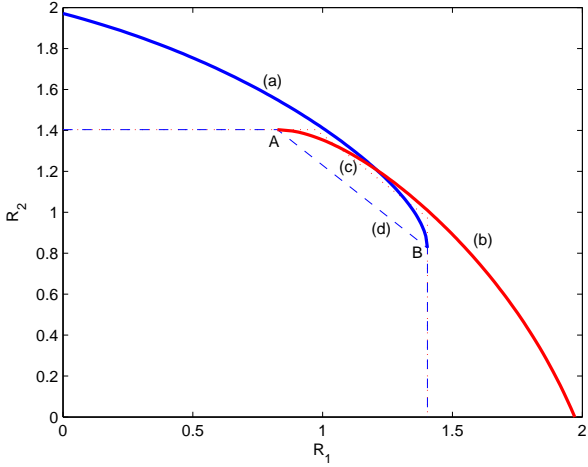}
\caption{The outer bounds of two-user Gaussian interference channel with $P_1=P_2=6$, $a^2=b^2=0.3$: (a) the capacity region $\Cif^{T_1}$; (b) the capacity region $\Cif^{T_2}$; (c) Carleial's outer bound in \cite{Carleial:IT83outbound}; (d) Kramer's outer bound in \cite{Kramer:IT04outbound} (theorem 1).}
\label{fig:ifc_outbound}
\end{figure}

On the other hand, the capacity region of GIFC-DMS can be served as the outer bound for GIFC.
Since both $\Gif^{T_1}$ and $\Gif^{T_2}$ include the capacity region of GIFC, we have 
\begin{equation} \label{eq:T1_T2}
\Gif \subset \Gif^{T_1} \cap \Gif^{T_1}\,.
\end{equation}
In Figure~\ref{fig:ifc_outbound}, we compare some known outer bounds for two-user Gaussian IFC for the same setup. 
As we can see, Kramer's outer bound in \cite{Kramer:IT04outbound} meets our outer bound of \eqref{eq:T1_T2} at point $A$, $B$, and performs better than ours elsewhere. 
Our outer bound in \eqref{eq:T1_T2} gives a better bound than Carleial's when $\alpha$ and $\beta$ in \eqref{eq:cap_R1} \eqref{eq:cap_R2} \eqref{eq:cap_T2} is close to 1. 

Note the outer bounds \eqref{eq:sum_T2}, \eqref{eq:sum_T1}  are proved based on a genie-aided argument,or more precisely ``receiver genie-aided'', in which some genie provides additional channel output to one of receivers.
In contrast, our approach here can be viewed as ``transmitter genie-aided'' in the sense that some genie gives additional information of channel input to one of the transmitters.
We comment that the outer bound by \eqref{eq:T1_T2} is not as good as the one obtained in \cite{Kramer:IT04outbound}.
However, the bound in \eqref{eq:T1_T2} might be possibly improved when one transmitter only knows partial information of the other message instead of the full message: for example, in Figure~\ref{fig:IFC_DMS} transmitter $T_1$ only know $g(W_2)$ instead of knowing $W_2$, where $g(\cdot)$ is some function of $W_2$.

\bigskip

\section{proofs}\label{sec:proof}
\subsection{Proof of Theorem~\ref{thm:achievable}} \label{sec:Proof_achieve}
The coding scheme is a straightforward application of Gelfand-Pinsker coding. 

\emph{Code Generation:} Fix $p(u,x_2)$, first generate $2^{n R_2}$ independent codewords of length $n$ at random according to the distribution $\prod_{i=1}^n p(u_i, x_{2,i})$ for message $w_2 \in \{1, \ldots, 2^{n R_2}\}$.  
Then generate $2^{n I(V; Y)}-\epsilon$ i.i.d. sequences $v^n$ according to the distribution $\prod_{i=1}^n p(v_i)$,and distribute these sequences uniformly into $2^{n R_1}$ bins.
For each sequence $v^n$, let $i(v^n)$ denote the index of the bin containing $v^n$.

\emph{Encoding:} Encoder 2 transmits $X_2(w_2)$.  
With $w_1$ and $w_2$, Encoder 1 looks in bin $w_1$ for a sequence $V^n$ that  $(X_1^n, V^n(w_1), X_2^n(w_2), U^n(w_2))$ is jointly typical and sends $X_{1,i}$.
If no such a sequence, an error is declared. 
If the number of sequences in each bin is larger than $2^{I(U,X_2; V)}$, the probability of finding no such $U^n$ decreases to 0 as n goes to $+\infty$.   

\emph{Decoding:} Receiver 2 determines the unique $\hat{\hat{W}}_2$ such that $(U^n(\hat{\hat{W}}_2), X_2^n(\hat{\hat{W}}_2), Y_2^n)$ is jointly typical.
Receiver 1 looks for the unique $V^n$ such that $(V^n, Y_1^n)$ is jointly typical and estimate the message $\hat{W}_1$ as the index of the bin containing the obtained $V^n$.

At Receiver 1, if 
\begin{equation*}
R_1 \leq I(V; Y_1) - I(V; U, X_2) -\epsilon\,,
\end{equation*}
the probability of error of $w_1$ decreases exponentially to zero as $n \to \infty$.
The probability of error of $w_2$ at receiver 2 goes to zero as $n \to +\infty$ if $R_2 \leq I(U, X_2; Y_2)$.

\subsection{Proof of Theorem~\ref{thm:outer_general}} \label{sec:Proof_outer}
Theorem~\ref{thm:outer_general} can be proved by adapting Marton's BC outer bound.  
For a $(R_1,R_2, n, P_{e,1}^{(n)}, P_{e,2}^{(n)})$ code with decoding error  $P_{e,i}^{(n)} \to 0$ as $n \to +\infty$, we define the auxiliary random variable $U$,
\begin{equation}
U_i=(W_2, Y_1^{i-1}, Y_{2,i+1}^{n}, X_{2}^{i-1}, X_{2,i+1}^n)\,.
\end{equation}

Applying Fano's inequality \cite{ElementsInfoTheory} for each message $W_t$, $t=1,2$, we have
\begin{equation}\label{eq:Fano_B}
H(W_t|Y_t^{n}) \leq n R_t P_{e,t}^{(n)} + h(P_{e,t}^{(n)}) = n \epsilon_t^{(n)}\,, 
\end{equation}
where $\epsilon_t^{(n)} \to 0$ as $P_{e,t}^{(n)} \to 0$.
Moreover, because transmitter 2 has no information about the message $W_1$, $X_2^n$ is independent of $W_1$ or the following relation holds
\begin{equation}\label{eq:entropy_cond}
H(W_1|W_2, X_2^n)=H(W_1)\,.
\end{equation}

To prove the converse, we need following lemmas:
\begin{lem}[\cite{Csiszar_Korner}]\label{lem:chain}
For any random variable $T$, the following equality holds,
\begin{multline}
\sum_{i=1}^n I(Y_{2,i+1}^{n}; Y_{1,i}|Y_1^{i-1}, T) \\
=\sum_{i=1}^n I(Y_{1}^{i-1};Y_{2,i}|Y_{2,i+1}^n, T)\,.
\end{multline}
\end{lem}

First let prove the outer bound for $R_1$ in \eqref{eq:bound_general}.
\begin{align}
n R_1 & \leq I(W_1; Y_1^n|X_2^n) + n \epsilon_1^{(n)} \label{eq:A_R1_a} \\
& = \sum_{i=1}^n I(W_1; Y_{1,i}|Y_{1}^{i-1}, X_2^n) + n \epsilon_1^{(n)} \nonumber \\
& \leq \sum_{i=1}^n \Bigl(H(Y_{1,i}|X_{2,i})-H(Y_{1,i}|X_2^n, Y_{1}^{i-1},X_{1,i})\Bigr) \nonumber \\
& \quad\quad\quad+ n \epsilon_1^{(n)}\nonumber \\
& \leq \sum_{i=1}^n I(X_{1,i};Y_{1,i}|X_{2,i}) + n \epsilon_1^{(n)} \nonumber\,.
\end{align}

Next we prove the outer bound for $R_2$ in \eqref{eq:bound_general}:
\begin{align}
n R_2 & \leq I(W_2; Y_2^{n}) + n \epsilon_2^{(n)} \label{eq:A_R2_a}\\
& \leq \sum_{i=1}^n I(W_2; Y_{2,i}|Y_{2,i+1}^{n}) + n \epsilon_2^{(n)} \nonumber \\
& \leq \sum_{i=1}^n I(W_2, Y_{2,i+1}^n; Y_{2,i}) + n \epsilon_2^{(n)}\nonumber\\
& \leq \sum_{i=1}^n I(U_i, X_{2,i}; Y_{2,i}) + n \epsilon_2^{(n)}\,, \nonumber
\end{align}
where \eqref{eq:A_R1_a} and \eqref{eq:A_R2_a} are from Fano's inequality in \eqref{eq:Fano_B}

Then we prove the outer bound for the sum rate $R_1+R_2$ in \eqref{eq:bound_general}:
\begin{subequations}
\begin{align}
\lefteqn{n(R_1+R_2)} \nonumber\\
& \leq  I(W_1; Y_1^n|W_2, X_2^n) + I(W_2; Y_2^n) + n\epsilon_1^{(n)} + n \epsilon_2^{(n)} \label{eq:A_sum2_a}\\
& \leq \sum_{i=1}^n \Bigl(I(W_1;Y_{1,i}|W_2,X_2^n,Y_{1}^{i-1})+ I(W_2;Y_{2,i}|Y_{2,i+1}^n)\Bigr) \nonumber\\
& \quad\quad\quad +n \epsilon_1^{(n)}+n \epsilon_2^{(n)} \nonumber \\
& \leq \sum_{i=1}^n \Bigl( I(W_1;Y_{1,i}|W_2,Y_{1}^{i-1},Y_{2,i+1}^n,X_2^n) \nonumber\\
& \quad\quad +I(Y_{2,i+1}^n;Y_{1,i}|W_2,Y_1^{i-1},X_2^n) \nonumber \\
& \quad\quad +I(W_2,Y_{1}^{i-1},Y_{2,i+1}^n,X_2^n; Y_{2,i}) \nonumber \\
& \quad\quad +I(Y_1^{i-1};Y_{2,i}|W_2,Y_{2,i+1}^n,X_2^n) + \epsilon_1^{(n)} + \epsilon_2^{(n)}\Bigr) \label{eq:A_sum2_b}\\
& \leq \sum_{i=1}^n\Bigl(I(W_1;Y_{1,i}|U_i,X_{2,i})+I(U_i,X_{2,i};Y_{2,i})\Bigr) \nonumber\\
& \quad\quad\quad + n\epsilon_1^{(n)} +n \epsilon_2^{(n)} \label{eq:A_sum2_c}\\
& \leq  \sum_{i=1}^n\Bigl(I(X_1;Y_{1,i}|U_i,X_{2,i})+I(U_i,X_{2,i};Y_{2,i})\Bigr) \nonumber\\
& \quad\quad\quad+n\epsilon_1^{(n)} +n \epsilon_2^{(n)}\,. \label{eq:A_sum2_d}
\end{align}
\end{subequations}
Note \eqref{eq:A_sum2_a} is due to Fano's inequality in \eqref{eq:Fano_B} and the conditional entropy relation in \eqref{eq:entropy_cond};
the two terms in \eqref{eq:A_sum2_b} are equal due to Lemma~\ref{lem:chain}; \eqref{eq:A_sum2_c} is obtained by using the auxiliary random variable $U_i$ and \eqref{eq:A_sum2_d} is true because
$(W_1,U_i) \to (X_{1,i}, X_{2,i}) \to (Y_{1,i},Y_{2,i})$ forms a Markov chain in this order for all $1 \leq i \leq n$.

\subsection{Proof of Theorem~\ref{thm:innerB}}\label{sec:Proof_innerB} 
Here we provide a proof for Theorem~\ref{thm:innerB}, the achievability of $\RR_{*}$ for type B weak interference.

\emph{Code Generation:} Fix $p(u,x_2)$, generate $2^{n R_2}$ independent codewords of length $n$ at random according to the distribution $\prod_{i=1}^n p(u_i, x_{2,i})$ for message $w_2 \in \{1, \ldots, 2^{n R_2}\}$. 
For each codeword $(U^n(w_2), X_2^n(w_2))$, generate $2^{n R_1}$ independent codewords $X(w_1,w_2)$ according to $\prod_{i=1}^n p(x_1|u,x_2)$, 
where $w_1 \in \{1, \ldots, 2^{n R_1}\}$.

\emph{Encoding:} Encoder 2 transmits $X_2(w_2)$. 
Since encoder 1 knows both messages, it sends $X_1(w_1,w_2)$. 

\emph{Decoding:} Receiver 2 determines the unique $\hat{\hat{W}}_2$ such that $(U^n(\hat{\hat{W}}_2), X_2^n(\hat{\hat{W}}_2), Y_2)$ is jointly typical.
Receiver 1 looks for the unique $(\hat{W}_1, \hat{W}_2)$ such that $(X_1^n(\hat{W}_1, \hat{W}_2), X_2^n(\hat{W}_2), U^n(\hat{W}_2))$ is jointly typical.

It is easy to see that the probability of error at receiver 2 goes to zero as $n \to +\infty$ if $R_2 \leq I(U, X_2; Y_2)$.  
Receiver 1 can decode $W_2$ successfully as $n \to +\infty$ if $R_2 \leq I(U, X_2; Y_1)$. 
For type B weak interference with \eqref{eq:typeB}, 
\begin{equation*}
R_2 \leq I(U X_2; Y_2) \leq I(U X_2; Y_1)\,.
\end{equation*} 
Thus receiver 1 can decode $W_2$ as long as receiver 2 can do so. 
With the error probability of $W_2$ going to zero as $n \to +\infty$, the error probability of $W_1$ at receiver 1 goes to zero if $R_1 \leq I(X_1; Y_1|X_2, U)$.
The above analysis shows that both receivers can decode with total probability of error going to 0 if \eqref{eq:R_star} is satisfied.
Hence there exists a sequence of good codes with error probability going to 0.

\subsection{Proof of Theorem~\ref{thm:capC}} \label{sec:Proof_capC}
For the proof of Theorem~\ref{thm:outer_general}, the key is to identify the auxiliary random variable and utilize the definition of type B weak interference. 

For any $(R_1, R_2, n, P_{e,1}^{(n)}, P_{e,2}^{(n)})$ code with decoding error  $P_{e,i}^{(n)} \to 0$ as $n \to +\infty$,
to prove the outer bound in \eqref{eq:R_star} for $R_1$, we consider
\begin{subequations}
\begin{align}
n R_1 &= H(W_1) \nonumber\\
& \leq I(W_1; Y_1^n)-n \epsilon_1^{(n)} \label{eq:B_R1_a} \\
&\leq I(W_1; Y_1^n | W_2, X_2^n) -n \epsilon_1^{(n)} \label{eq:B_R1_b}\\
& \leq  \sum_{i=1}^n I(W_1; Y_{1,i}|W_2, Y_1^{i-1},X_2^{n}) -n \epsilon_1^{(n)}\label{eq:B_R1_c}
\end{align}
where the inequality \eqref{eq:B_R1_a} is due to Fano's inequality \eqref{eq:Fano_B}; \eqref{eq:B_R1_b} follows from \eqref{eq:entropy_cond} and \eqref{eq:B_R1_c} is due to the reason that the mutual information will be increases by adding conditionals. 
Define $U_i=(W_2, Y_1^{i-1}, X_2^{i-1})$,
\begin{align}
n R_1 & \leq \sum_{i=1}^n I(W_1; Y_{1, i}|U_i, X_{2,i}, X_{2,i+1}^n) +n \epsilon_1^{(n)} \nonumber\\
& = \sum_{i=1}^n \Bigl(H(Y_{1,i}|U_i, X_{2,i}, X_{2,i+1}^n) \nonumber\\
& \quad\quad - H(Y_{1,i}| U_i, X_{2,i}, X_{2,i+1}^n, W_1)\Bigr) +n \epsilon_1^{(n)}  \label{eq:B_R1_d}\\
& \leq \sum_{i=1}^n \Bigl(H(Y_{1,i}|U_i, X_{2,i}) \nonumber \\
& \quad- H(Y_{1,i}|U_i,X_{2,i},X_{2,i+1}^n,W_1,X_{1,i})+\epsilon_1^{(n)}\Bigr) \label{eq:B_R1_e} \\
& =\sum_{i=1}^n \Bigl( H(Y_{1,i}|U_i,X_{2,i})  \nonumber \\
& \quad\quad - H(Y_{1,i}|U_i,X_{1,i},X_{2,i}) + \epsilon_1^{(n)}\Bigr)\label{eq:B_R1_f} \\
& =\sum_{i=1}^n \Bigl(I(X_{1,i}; Y_{1,i}|U_i,X_{2,i}) + \epsilon_1^{(n)}\Bigr)\label{eq:B_R1_g}
\end{align}
\end{subequations}
\eqref{eq:B_R1_e} is because the entropy increases when dropping some conditionals and it decreases by adding more conditions.
The equality \eqref{eq:B_R1_f} is true because $W_1-(X_{1,i}, X_{2,i})-Y_{1,i}$ forms a Markov chain. 
 
Now for the bound of $R_2$ in \eqref{eq:R_star}, we have
\begin{subequations}
\begin{align}
n R_2 &=H(W_2) \nonumber\\
& \leq I(W_2; Y_2^n) + n\epsilon_2^{(n)} \label{eq:B_R2_1}\\
& = \sum_{i=1}^n I(W_2; Y_{2,i}|Y_2^{i-1})+ n\epsilon_2^{(n)} \nonumber \\
& \leq \sum_{i=1}^n \Bigl( H(Y_{2,i})-H(Y_{2,i}|Y_2^{i-1},W_2, X_2^{i}) + \epsilon_2^{(n)}\Bigr)\,, \label{eq:B_R2_2}
\end{align}
\end{subequations}
where \eqref{eq:B_R2_1} follows from Fano's inequality in \eqref{eq:Fano_B}; 
\eqref{eq:B_R2_2} is because the entropy will decrease when adding conditionals.

Note for the interference channel two receivers cannot cooperate, the capacity region is the same as the one with the same marginal output $p(y_1|x_1,x_2)$, $p(y_2|x_1,x_2)$. 
Thus for type B weak interference satisfying \eqref{eq:typeB}, given $X_2$, one can construct a Markov chain, namely,
\begin{equation*}
X_1 \to Y_1 \to Y_2\,.
\end{equation*}
Thus the conditional entropy in \eqref{eq:B_R2_2} satisfies 
\begin{subequations}
\begin{align}
\lefteqn{H(Y_{2,i}|Y_2^{i-1},W_2, X_2^i)} \nonumber\\
& \geq H(Y_{2,i}|Y_1^{i-1},W_2, X_2^{i}) \label{eq:B_R2_a}\\
& =H(Y_{2,i}|U_i,X_{2,i}) \label{eq:B_R2_b}
\end{align}
\end{subequations}
Combine \eqref{eq:innerC} and \eqref{eq:B_R2_2}, \eqref{eq:B_R2_b}, we get the result in Theorem~\ref{thm:capC}. 

\subsection{Proof of Theorem~\ref{thm:GIFC}}
Achievability: The achievability of this rate utilizes the now famous dirty-paper coding strategy. First, we generate a codebook of $2^{nR_2}$ codewords according to ${\mathcal{N}}(0,\Sigma)$, where $\Sigma$ is the covariance between transmitter 1 and 2. Transmitter 1 devotes $(1- \alpha)$ fraction of its power $P_1$ to the transmission of $W_2$, while Transmitter 2 devotes its entire power $P_2$ to this effort. This leads to a covariance of the form

\begin{equation}
\Sigma = \left[\begin{array}{cc} (1-\alpha)P_1 & \gamma \\ \gamma & P_2 \end{array}\right]
\end{equation}
 
The effective interference seen by Receiver 1 is a combination of the signals communicated from both Transmitters 1 and 2. Since Transmitter 1 knows the exact realization of the message $w_2 \in W_2$, it has non-causal side information on the interference and can completely cancel it out, achieving a rate
\begin{equation}
R_1 = \frac{1}{2} \log\left(1+ \alpha P_1 \right)
\end{equation}
using a Gaussian codebook with codewords that are correlated with the interference. At Receiver 2, this Gaussian codebook for $W_1$ is perceived as additive interference, hence achieving a rate:

\begin{equation}
R_2= \frac{1}{2} \log\left(1 + \frac{h \Sigma h^t}{1+ b^2 \alpha P_1}\right)
\end{equation}
Maximize $R_2$ over $\abs{\gamma}^2 \leq (1-\alpha)P_1 P_2$ (such that $\Sigma$ is positive definite), it is not difficult to shown $R_2$ obtain the maximum when $\gamma = \sqrt{(1-\alpha)P_1 P_2}$ and \eqref{eq:cap_R2} can be achieved. 

\bigskip
Converse:
Since Gaussian IFCs with degraded message sets and $\abs{b} \leq 1$ satisfy the condition \eqref{eq:typeC} for type C weak interference, the outer bound \eqref{eq:R_star} holds. 
The central feature here is to prove the optimality of Gaussian input. 

To prove the Gaussian optimality, we need following lemmas:
\begin{lem}[Lemma 1 in \cite{Thomas:feedback_dblcap}] \label{lem:thomas_bnd}
Let $X_1, X_2, \ldots, X_k$ be an arbitrary set of zero-mean random variables with covariance matrix $K$. Let $S$ by any subset of $\{1,2,\ldots,k\}$ and $\bar{S}$ be its complement. Then
\begin{equation*}
h(X_{S}|X_{\bar{S}}) \leq h(X_{S}^*|X_{\bar{S}}^*)\,,
\end{equation*} 
where 
\begin{equation*}
(X_1^*, X_2^*, \ldots, X_k^*) \thicksim N(0,K)\,.
\end{equation*}
\end{lem}

\begin{lem}\label{lem:cond_exp}
Let $X_1, X_2$ be arbitrarily distributed zero-mean random variables and $X_1^*, X_2^*$ be zero-mean Gaussian distributed random variables with the same covariance matrix as $X_1, X_2$, then 
\begin{equation*}
\Exp[X_1 X_2] \leq \Bigl(\Exp\bigl[(\Exp[X_1^*|X_2^*])^2\bigr] \Exp\bigl[ (X_2^*)^2\bigr]\Bigr)^{\frac{1}{2}} 
\end{equation*}
\end{lem}
\begin{proof}
Using Cauchy inequality and the properties of conditional expectation,
\begin{align*}
\Exp[X_1 X_2] & =\Exp[X_1^* X_2^*] \\
& = \Exp\bigl[\Exp[X_1^* X_2^*|X_2^*]\bigr] \\
& = \Exp\bigl[\Exp[X_1^*|X_2^*]X_2^*\bigr] \\
& \leq \Bigl(\Exp\bigl[(\Exp[X_1^*|X_2^*])^2\bigr] \Exp\bigl[ (X_2^*)^2\bigr]\Bigr)^{\frac{1}{2}} \,.
\end{align*}
\end{proof}

Let denote $X_1^*,X_2^*, U^*$ are Gaussian distributed random variables with the same covariance with $X_1, X_2,U$, and $\var(A|B)$ denotes the conditional variance of $A$ given $B$, namely,
\begin{equation*}
\var{A|B}=\Exp[A^2|B]\,.
\end{equation*} 
First we note that 
\begin{align*}
0 & \leq h(Y_1|U, X_2) \\
& = h(X_1 + a X_2 + Z_1| U, X_2) \\
& = h(X_1+Z_1|U, X_2) \\
& \leq h(X_1+Z_1) \\
& \leq \frac{1}{2} \log(1+P_1)\,.
\end{align*}  
Without losing any generality, we can assume 
\begin{equation} \label{eq:R1_entropy}
h(Y_1|U, X_2) = \frac{1}{2}\log(1+ \alpha P_1)\,, 
\end{equation} 
for some $\alpha$, $0 \leq \alpha \leq 1$. Thus we have 
\begin{align}
I(X_1; Y_1| U, X_2) & = h(Y_1|U, X_2)-h(Y_1|U,X_1,X_2) \nonumber \\
& = \frac{1}{2}\log(1+ \alpha P_1) \label{eq:R1_proof}
\end{align}
On the other hand, by Lemma~\ref{lem:thomas_bnd},
the conditional entropy is upper bounded by the Gaussian variables with the same covariance matrix, thus,
\begin{eqnarray*}
h(Y_1|U, X_2) & \leq & h(X_1+Z_1|X_2) \\ 
& \leq & h(X_1^*+Z_1|X_2^*) \\ 
& = & \frac{1}{2}\log\bigl(1+\var{X_1^*|X_2^*}\bigr)\,.
\end{eqnarray*}
Combining with \eqref{eq:R1_entropy}, we have a lower bound for $\var{X_1^*|X_2^*}$, 
\begin{equation}\label{eq:cond_bnd}
\var{X_1^*|X_2^*} \ge \alpha P_1\,.
\end{equation} 
Note 
\begin{equation*}
\var{X_1^*|X_2^*}=\Exp[(X_1^*)^2]-\Exp\bigl[(\Exp[X_1^*|X_2^*])^2\bigr]\,,
\end{equation*} 
together with \eqref{eq:cond_bnd}, we have
\begin{equation} \label{eq:var_bnd}
\Exp\bigl[(\Exp[X_1^*|X_2^*])^2\bigr] \leq (1-\alpha) P_1\,.
\end{equation} 
Combine Lemma~\ref{lem:cond_exp} and \eqref{eq:var_bnd},    
\begin{equation*}
\Exp[X_1 X_2] \leq \sqrt{(1-\alpha) P_1 P_2} \,.
\end{equation*}
Thus 
\begin{align}
h(Y_2) &= h(b X_1+X_2+Z_2) \nonumber\\
& \leq \frac{1}{2}\log(1+b^2 P_1 + P_2 + 2b \Exp[X_1 X_2]) \nonumber\\
& \leq \frac{1}{2}\log(1+b^2 P_1 + P_2 + 2b\sqrt{(1-\alpha)P_1 P_2})\,. \label{eq:Y2_bnd}
\end{align}
At last we need to bound $h(Y_2|X_2, U)$. 
Define $Y_1', Y_2'$ as follows, 
\begin{align*}
Y_1' = X_1 + Z_1 \\
Y_2' = b X_1+Z_2 \,.
\end{align*}
Thus $Y_2'$ is a stochastically degraded version of $Y_1'$, or in other words, $Y_2'=b Y_1'+ Z'$, where $Z'$ is Gaussian distributed with variance $1-b^2$.
By entropy power inequality (EPI) \cite{ElementsInfoTheory},
\begin{align*}
2^{2 h(Y_2|X_2, U)} & = 2^{2 h(Y_2'|X_2,U)} \\
& \ge 2^{2 h(b Y_1'|X_2, U)}+2^{2 h(Z')} \\
& = b^2 2^{2 h(Y_1|X_2, U)}+ 1-b^2 \\
& = 1+b^2 \alpha P_1 \,,
\end{align*} 
thus 
\begin{equation}\label{eq:Y2_cond_bnd}
h(Y_2|X_2, U) \ge \frac{1}{2}\log(1+b^2 \alpha P_1)\,.
\end{equation} 
And finally we combine \eqref{eq:Y2_bnd} and \eqref{eq:Y2_cond_bnd},
\begin{align}
\lefteqn{I(X_2, U; Y_2)} \nonumber\\
& = h(Y_2) - h(Y_2|X_2,U) \nonumber \\
& \leq \frac{1}{2}\log\bigl(\frac{1+b^2 P_1 + P_2 + 2b\sqrt{(1-\alpha)P_1 P_2}}{1+b^2 \alpha P_1}\bigr)\,.\label{eq:R2_proof}
\end{align}
Since \eqref{eq:R1_proof} \eqref{eq:R1_proof} are similar to \eqref{eq:cap_R1} \eqref{eq:cap_R2}, the optimality of Gaussian inputs is established and the proof for converse is complete. 

\section{Conclusions and Future Work}\label{sec:conclude}
In this paper, we investigate the capacity region of two-user interference channel (IFC) with degraded message sets (DMS), in which one transmitter knows both messages. 
For the general discrete memoryless interference channel setting, we find an achievable region and outer bound, which meet under a weak interference condition. Thus, we determine the capacity region of a class of channels that includes the Gaussian weak interference channel.

Possible extensions to this work include: 
\begin{enumerate}
\renewcommand{\theenumi}{\roman{enumi}}
\item Lossy functions of the message ($w_2$) made available to $T_1$ rather than the message itself. This might possibly yield a better outer bound for the non-cooperative IFC. 
\item Generalizing this approach to IFC with more than two users in the system.
\end{enumerate}
\section*{Acknowledgments}
This research was supported in part by the Office of Naval 
Research through the Electric Ship Research and Development
Consortium, in part by the National Science Foundation under Grants
ECS-0218207 and ECS-0225448.
Wei Wu was also supported by the Hemphill-Gilmore student Endowed Fellowship through the University of Texas at Austin. 

\bibliographystyle{IEEEtran}
\bibliography{ifc_nfb}
\end{document}